\documentclass[pdflatex,sn-mathphys-num]{sn-jnl}% Math and Physical Sciences Numbered Reference Style
\RequirePackage{fix-cm}
\usepackage{graphicx}
\usepackage{hyperref}
\usepackage{multirow}
\usepackage{multicol}
\usepackage{array}
\usepackage{amsmath,amssymb,amsfonts}
\usepackage{amsthm}
\usepackage{mathrsfs}
\usepackage[title]{appendix}
\usepackage{xcolor}
\usepackage{textcomp}
\usepackage{manyfoot}
\usepackage{rotating}
\usepackage{tabularx}
\usepackage{booktabs}
\usepackage{float}
\usepackage{subcaption}
\usepackage{algorithm}
\usepackage{algorithmicx}
\usepackage{algpseudocode}
\usepackage{listings}
\usepackage{pifont}
\usepackage{makecell}
\newcommand{\cmark}{\ding{51}} % Tick
\newcommand{\xmark}{\ding{55}} % Cross
\theoremstyle{thmstyleone}%
\theoremstyle{thmstyletwo}%

\theoremstyle{thmstylethree}%

\begin{document}

\title[Article Title]{Enhancing Software Vulnerability Detection Through Adaptive Test Input Generation Using Genetic Algorithm}


%\author*[1]{\fnm{Yanusha} \sur{Mehendran}}\email{yanusha.mehendran@hdr.qut.edu.au}
\author[1]{\fnm{Yanusha} \sur{Mehendran}}\email{yanusha.mehendran@hdr.qut.edu.au}

\author[1]{\fnm{Maolin} \sur{Tang}}\email{m.tang@qut.edu.au}

\author[1]{\fnm{Yi} \sur{Lu}}\email{yt.lu@qut.edu.au}

\affil[1]{\orgdiv{School of Computer Science}, \orgname{Queensland University of Technology}, \orgaddress{\city{Brisbane}, \postcode{4000}, \state{QLD}, \country{Australia}}}

%\affil[2]{\orgdiv{Department}, \orgname{Organization}, \orgaddress{\street{Street}, \city{City}, \postcode{10587}, \state{State}, \country{Country}}}

%\affil[3]{\orgdiv{Department}, \orgname{Organization}, \orgaddress{\street{Street}, \city{City}, \postcode{610101}, \state{State}, \country{Country}}}


\abstract{Software vulnerabilities continue to undermine the reliability and security of modern systems, particularly as software complexity outpaces the capabilities of traditional detection methods. This study introduces a genetic algorithm-based method for test input generation that innovatively integrates genetic operators and adaptive learning to enhance software vulnerability detection. A key contribution is the application of the crossover operator, which facilitates exploration by searching across a broader space of potential test inputs. Complementing this, an adaptive feedback mechanism continuously learns from the system’s execution behavior and dynamically guides input generation toward promising areas of the input space.
Rather than relying on fixed or randomly selected inputs, the approach evolves a population of structurally valid test cases using feedback-driven selection, enabling deeper and more effective code traversal. This strategic integration of exploration and exploitation ensures that both diverse and targeted test inputs are developed over time.
Evaluation was conducted across nine open-source JSON-processing libraries. The proposed method achieved substantial improvements in coverage compared to a benchmark evolutionary fuzzing method, with average gains of 39.8\% in class coverage, 62.4\% in method coverage, 105.0\% in line coverage, 114.0\% in instruction coverage, and 166.0\% in branch coverage. These results highlight the method’s capacity to detect deeper and more complex vulnerabilities, offering a scalable and adaptive solution to software security testing.}

\keywords{Software Testing, Evolutionary Algorithms, Test Input Generation, Genetic Operators, Grammar-Based Fuzzing}



\maketitle

\section{Introduction}\label{sec1}

Software is one of the essential elements of any computer system, being programs and applications that help the hardware perform a specific task. However, with the growing complexity of software, there are vulnerabilities which may lead to the loss of the system’s integrity, confidentiality, or availability. A software vulnerability may arise from coding errors, design flaws, or misconfigurations, each posing significant risks to digital security. Ideally, software should be written in a manner that eliminates vulnerabilities entirely. However, achieving this level of flawlessness is practically impossible due to factors such as the complexity of modern systems, the inevitability of human error, and the dynamic nature of evolving threats. As a result, no system can be entirely free from vulnerabilities, underscoring the necessity of robust vulnerability detection techniques to mitigate potential risks \cite{godefroid2012sage}.

Traditional manual approaches to vulnerability detection face several challenges. These methods rely heavily on human expertise for feature extraction \cite{guo2020vulhunter,li2018vuldeepecker}, making them susceptible to the limitations of human capacity and consistency. Large amounts of code that needs to be analyzed may overwhelm experts and vulnerabilities may be missed. Also, manual analysis can introduce subjectivity into the assessment process. Manual analysis is labor intensive and time consuming, which requires an in-depth code review that may introduce human errors \cite{8769937,li2017mining,niu2020deep}. Not only does this delay the detection of vulnerabilities but it also leaves the system exposed to vulnerabilities for a longer period of time, highlighting the
importance of automated and fast techniques such as fuzzing.   

Among the various methods for identifying vulnerabilities, fuzzing has emerged as a particularly effective approach \cite{godefroid2020fuzzing}. Fuzzing involves generating diverse and often unexpected inputs to test software systems, systematically exploring execution paths to uncover defects. Fuzzing itself encompasses several approaches, including static, dynamic, and hybrid methods \cite{Shin_Williams_2011a}. Static fuzzing generates test inputs without executing the software, relying on predefined rules or formats to uncover potential vulnerabilities. While useful for quick validation, it is prone to high false positive rates and often lacks the context needed to uncover dynamic, runtime-specific vulnerabilities. Dynamic fuzzing, on the other hand, executes the software and monitors its behavior in real-time, making it effective for identifying vulnerabilities that arise under specific runtime conditions \cite{li2018fuzzing}. Hybrid fuzzing combines elements of both static and dynamic methods, offering a more comprehensive testing approach by leveraging the strengths of each. Despite these advancements, a persistent challenge in fuzzing remains: generating high-quality and diverse test inputs. Traditional random input generation methods often fail to explore complex software paths effectively. The scarcity of effective test inputs further intensifies the difficulty of discovering vulnerabilities in increasingly complex systems, highlighting the need for innovative techniques to address this gap.

In response to these challenges, this paper introduces a novel evolutionary computation-based approach for optimizing test input generation in dynamic fuzzing. Evolutionary computation, inspired by the principles of natural selection, iteratively refines a population of test inputs through operations such as mutation, crossover, and selection \cite{veggalam2016ifuzzer}. This adaptive process enables the generation of test inputs that are not only diverse but also highly effective in uncovering vulnerabilities. Among the class of evolutionary computation methods, this study specifically adopts a genetic algorithm (GA), a well-established method that evolves candidate solutions by mimicking biological evolution. Genetic algorithms offer a well-structured way to introduce variation and apply selection, making them especially effective for exploring complex input spaces in fuzz testing. By leveraging feedback from the Software Under Test (SUT), the genetic algorithm dynamically improves input quality, ensuring broad exploration and targeted refinement.

The proposed GA approach aims to overcome the limitations of traditional input generation methods by introducing adaptability, generalizability, and structural awareness into the fuzzing process. Traditional fuzzers often rely on seed inputs or domain-specific heuristics, which can constrain the diversity and depth of generated test inputs. In contrast, this study presents a \textbf{novel grammar-guided genetic algorithm (GA)} for dynamic fuzzing that begins from scratch which is a cold-start setting and generates valid structured inputs without relying on existing test cases or prior knowledge of the SUT.

This study contributes to the domain of software vulnerability detection through the development of a grammar-driven, domain-agnostic GA based fuzzing framework. Evolutionary computation encompasses a variety of algorithms inspired by natural evolution, including Genetic Algorithms (GA), Genetic Programming (GP), Evolution Strategies (ES), and Differential Evolution (DE). While all these approaches share the core principle of evolving candidate solutions over generations, they differ primarily in how individuals are represented and manipulated. In Genetic Programming, solutions are represented as executable program trees, making it suitable for evolving logic structures. In contrast, Genetic Algorithms operate on structured or fixed-length representations, making them particularly well-suited for problems like test input generation, where candidate inputs such as JSON files can be directly encoded and manipulated.

This study adopts a GA because of its strength in evolving structured representations with controlled variation. The initial population of JSON test inputs is generated directly from a JSON grammar, ensuring syntactic validity while offering diversity. Notably, this cold-start generation approach eliminates the dependency on initial seeds, making the method applicable to diverse grammar-defined input format. From this base, the algorithm evolves inputs using genetic operators specifically tailored for structured data. The mutation operator introduces localized structural changes, while the crossover operator recombines substructures from two parent inputs, enabling the discovery of novel and complex structural patterns. This grammar-aware crossover is particularly powerful, as it preserves syntactic validity while expanding the input space far beyond what mutation alone can achieve.

The evolutionary process is adaptive and feedback-driven where the generated inputs are executed against target SUTs, and branch coverage and exceptions are collected as feedback to guide the fitness evaluation. This optimization process enables the algorithm to dynamically target unexplored code regions, improving the likelihood of uncovering subtle or deep vulnerabilities that traditional approaches might miss.

The novelty of this GA approach lies in its grammar-guided, cold-start fuzzing approach that eliminates the reliance on manually crafted seed inputs or domain-specific heuristics. By generating test inputs directly from a predefined grammar, the method is inherently domain-agnostic and adaptable to a wide range of structured formats. Furthermore, the study introduces a structure-preserving crossover and mutation operator that recombines syntactically valid substructures from parent inputs, allowing the exploration of complex and diverse input patterns while maintaining grammatical correctness. In addition, the approach adopts a feedback-driven optimization strategy based solely on branch coverage, simplifying the fitness evaluation process and avoiding the complexity of multi-objective tuning commonly seen in prior work. The key contributions of this study are as follows.
\begin{enumerate}
    \item the design and implementation of a novel grammar-based GA that applies both structure-preserving crossover and mutation operators to generate valid, diverse JSON inputs from scratch.
    \item development of a feedback-driven fuzzing loop that dynamically evolves the input population based on execution feedback from the SUTs
    \item comprehensive evaluation on real-world software systems that require structured inputs, demonstrating the approach’s broad applicability and its ability to uncover new execution paths
\end{enumerate}

\section{Related Work}\label{sec2}

The field of software vulnerability detection has seen significant advancements through the development of various testing methodologies and optimization techniques. Among these, fuzzing has emerged as a prominent strategy for uncovering software defects, while evolutionary computation has demonstrated its effectiveness in enhancing automated test case generation.

\subsection{Fuzzing Approaches}\label{subsec1}

Fuzzing has emerged as a powerful technique for uncovering software vulnerabilities by generating and executing diverse test inputs to explore potential defects. Over the years, various fuzzing approaches have been developed, each tailored to address specific challenges in software testing. These approaches can be broadly categorized based on their input generation strategies, level of access to the software, and testing methodologies. This section examines static, dynamic, and hybrid fuzzing approaches, providing an analysis of their methodologies, strengths, and limitations.

Static analysis provides high detection speed, enabling quick examination and issue resolution. However, static analysis often suffers from high false positive rates, as many tools lack user-friendly vulnerability detection models, leading to numerous false alarms. This complicates the identification and validation of actual vulnerabilities \cite{li2018fuzzing,liu2008vulnerability}. Zheng et al. introduced a static fuzzy mutation method guided by software vulnerability evolution laws, identifying potential threat paths through Abstract Syntax Tree (AST) analysis \cite{zheng2023abstract}. In a related direction, Ponta et al. \cite{ponta2020detection} proposed a comprehensive approach combining static reachability analysis with dynamic reachability assessment to detect, assess, and mitigate vulnerabilities in open source dependencies.

While dynamic analysis (e.g., fuzzing) can achieve high precision in vulnerability detection, it often requires significant manual effort for debugging and triaging results, demanding technical expertise \cite{li2018fuzzing}. Additionally, dynamic approaches face scalability challenges due to runtime overhead. Dynamic analysis underpins techniques like black-box, white-box, and grey-box fuzzing, which optimize the exploration of software vulnerabilities. Recent advances in grey-box fuzzing focus on intelligent mutation strategies, such as CMFuzz \cite{wang2021cmfuzz}, which employs contextual bandit algorithms (LinUCB) to dynamically select optimal mutation operators based on seed file characteristics. This context-aware approach significantly improves code coverage and crash discovery compared to traditional uniform mutation strategies. The selection of initial seed inputs has also been shown to significantly affect fuzzing effectiveness, as observed in a study on JavaScript engines where seeds derived from CVE proof-of-concepts led to higher code coverage and more crashes \cite{wen2023evaluating}. STATEAFL further advances grey-box fuzzing by targeting stateful network servers, which require multiple sequential inputs to reach deep protocol states \cite{natella2022stateafl}. 

Hybrid fuzzing integrates the strengths of fuzzing and symbolic execution to overcome their respective limitations, offering a robust framework for effective vulnerability detection. Kim et al. \cite{kim2020hfl} introduced HFL, a hybrid approach tailored for kernel fuzzing, which incorporates mechanisms for syscall sequence inference and resolving nested argument structures, significantly enhancing code coverage and vulnerability detection in Linux kernels. GAFuzzing combines static and dynamic techniques to improve traditional fuzzing by extracting structural information from code to guide test generation \cite{liu2008vulnerability}. Yun et al. \cite{yun2018qsym} proposed QSYM, a practical concolic execution engine that applies symbolic execution selectively to challenging paths identified during fuzzing, thereby optimizing efficiency and scalability while achieving deeper code path exploration. He et al. \cite{he2020toward} extended hybrid fuzzing to IoT firmware, integrating static analysis for known vulnerabilities and dynamic fuzzing for discovering new ones, demonstrating its effectiveness in addressing IoT-specific security challenges. Together, these studies highlight the versatility and efficacy of hybrid fuzzing in uncovering complex vulnerabilities across diverse domains, including operating systems, IoT devices, and general software systems. Zhang et al. proposed a coverage-guided fuzzing method that employs reinforcement learning-enabled multi-level input mutation to dynamically adapt the mutation strategy based on feedback, significantly improving vulnerability discovery efficiency. Recent comprehensive reviews comprehensive surveys have explored the application of deep learning techniques in identifying software vulnerabilities \cite{pham2024coverage}. Zhu et al. extensively reviewed 48 DL-based studies, emphasizing the perception gap between machine-learned models and expert-level vulnerability understanding. Their study highlights both the potential and the current limitations of deep learning techniques in addressing complex vulnerability semantics \cite{zhu2023application}.

\subsection{Evolutionary Computation in Fuzzing}\label{subsec2}

Evolutionary computation, inspired by the principles of natural selection, have found applications across diverse fields, including software engineering. These algorithms have been widely studied and adapted to tackle complex problems by iteratively optimizing solutions. In the context of software testing, evolutionary computation play a pivotal role in automating the generation of test cases. Their adaptability makes them particularly effective in addressing the dynamic challenges of modern software systems, including fuzzing, where generating diverse and high-quality inputs is crucial for detecting vulnerabilities.

The target domains of evolutionary fuzzing approaches are diverse, depending on the SUT. The study on Vulnerability Analysis for X86 Executables focuses on optimizing input generation for x86 executables using genetic algorithms \cite{liu2008vulnerability}. VUzzer addresses binary applications by leveraging application-aware feedback to guide the fuzzing process \cite{rawat2017vuzzer}. In the domain of web applications, KameleonFuzz is specifically tailored to uncover cross-site scripting (XSS) vulnerabilities \cite{duchene2014kameleonfuzz}. Additionally, IFuzzer and EvoGFuzz target structured input domains, refining test inputs for interpreters and leveraging grammar-based approaches \cite{veggalam2016ifuzzer,eberlein2020evolutionary}.

The input structures used in these studies align with their respective target domains. Binary fuzzers primarily handle inputs without predefined structure, enabling them to test a wide range of applications \cite{rawat2017vuzzer}. In contrast, methods targeting web applications often use attack grammars to craft inputs tailored to specific vulnerabilities, focusing on structured contexts \cite{duchene2014kameleonfuzz}. Grammar-based input generation has been effective in producing syntactically valid structured inputs for domains requiring strict adherence to input formats \cite{eberlein2020evolutionary}. However, despite dealing with structured inputs, EvoGFuzz does not employ a fully-fledged genetic algorithm (GA) since it lacks crossover as a genetic operator, a key component of traditional evolutionary algorithms.

The genetic operators utilized highlight variations in evolutionary strategies. VUzzer primarily relies on mutation as its dominant operator, guided by dynamic taint analysis and control-flow features to mutate input bytes at critical offsets \cite{rawat2017vuzzer}. Similarly, KameleonFuzz emphasizes mutation over crossover, using grammar-guided mutations to refine XSS payloads while limiting crossover to subtree swaps. IFuzzer employs GP that uses crossover and mutation to evolve test inputs, focusing on interpreter-specific requirements \cite{veggalam2016ifuzzer}. In contrast, EvoGFuzz does not employ crossover, relying entirely on mutation to evolve inputs generated from a probabilistic grammar \cite{eberlein2020evolutionary}. This design was chosen because directly mutating individuals may lead to syntactically invalid inputs that violate the grammar, and the probabilistic grammar itself already introduces stochasticity into the generation process. 

The fitness functions used in these studies are tailored to their respective objectives. KameleonFuzz relies on feedback from successful XSS payloads to optimize input generation \cite{duchene2014kameleonfuzz}. EvoGFuzz \cite{eberlein2020evolutionary} and IFuzzer \cite{veggalam2016ifuzzer} use fitness functions based on execution behavior, such as triggered exceptions and structural complexity. Specifically, EvoGFuzz evaluates fitness through a multi-objective function that combines feedback scores, which measure the ability of inputs to trigger exceptions, and structural complexity, which assesses adherence to the probabilistic grammar. Similarly, IFuzzer employs a fitness function that evaluates structural complexity metrics, including cyclomatic complexity, to guide the generation of inputs that explore intricate execution paths. Studies like \cite{eberlein2020evolutionary} and \cite{veggalam2016ifuzzer} employ fitness functions to address multiple objectives, though they do not appear to leverage weight-based approaches, which could enhance the balancing of diverse goals during optimization.

The motivation for this study arises from the inherent challenges in generating effective and diverse structured inputs for software vulnerability detection. Structured inputs, such as JSON, require strict syntactic correctness while also demanding extensive exploration of the input space to expose hidden vulnerabilities. Existing genetic algorithm-based fuzzing techniques have largely focused on mutation operators. While mutation supports fine-grained local search, it may struggle to escape limited regions of the input space. In contrast, this study emphasizes the integration of crossover that fundamentally recombines segments from different parent inputs to introduce entirely new structural variations that would be unlikely to emerge through mutation alone. Crossover enables the exploration of a much broader and more diverse input space, increasing the likelihood of triggering deeper execution paths and revealing complex vulnerabilities that traditional mutation-based fuzzing may miss. Crucially, both crossover and mutation operators are carefully designed to preserve the syntactic validity of the generated inputs.

Furthermore, the approach adopts an adaptive, feedback-driven learning mechanism that continuously refines input generation based on runtime behavior, allowing the algorithm to dynamically prioritize under-tested code regions. This dual focus on structural diversity through genetic operators and adaptive exploration through feedback distinguishes the proposed method from existing GA-based approaches, offering a more robust framework for advancing software vulnerability detection. Furthermore, its grammar-guided, seed-independent design making it a versatile solution for diverse testing scenarios.

\section{Design of the New GA-based Approach}\label{sec3}

Classic input generation methods face challenges in effectively exploring complex and structured input spaces, often leaving critical execution paths untested. Additionally, the scarcity of high-quality and diverse test data further exacerbates the difficulty in achieving comprehensive software testing, limiting the discovery of potential vulnerabilities. To address this, our approach utilizes a Genetic Algorithm (GA) as its foundation, leveraging its capability for dynamic evolution and optimization. By iteratively refining a population of candidate inputs, the algorithm generates inputs that maintain structural validity while adapting to the feedback from the SUT. This adaptive and iterative process ensures both diversity and effectiveness, enabling a more thorough and targeted vulnerability detection process.

To address the challenges of input diversity and dynamic exploration, the process begins with generating test inputs from a predefined grammar, ensuring that they adhere to the required structure for effective testing. This grammar defines the rules for generating syntactically correct inputs, enabling the creation of a random initial population of 100 inputs. These inputs form the foundation for the evolutionary process, providing a diverse set of test cases for subsequent analysis.

\begin{figure}[h]
    \centering
    \includegraphics[width=1\textwidth]{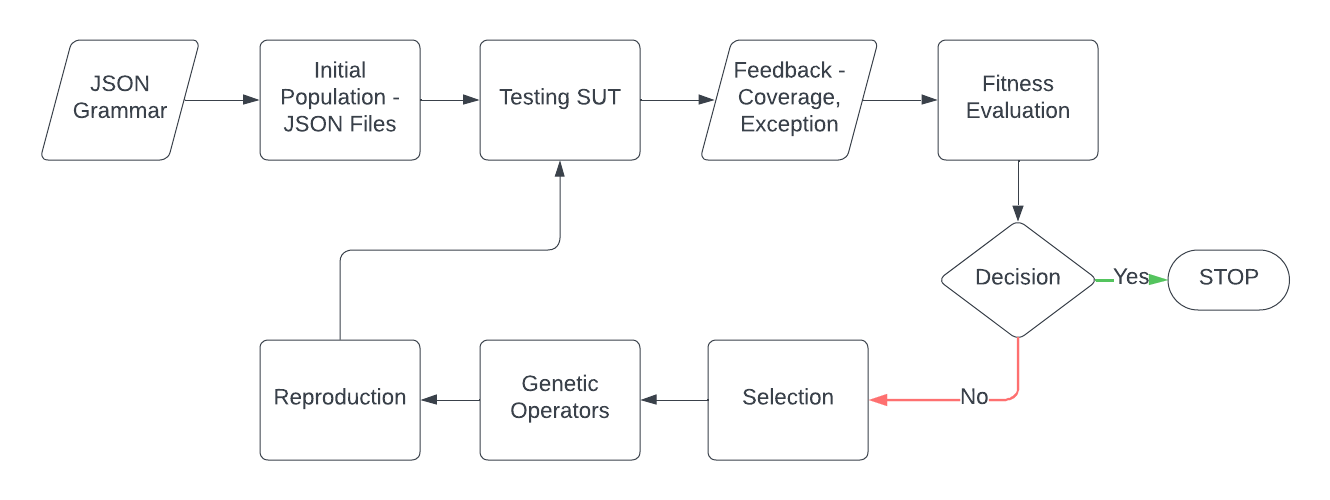}
    \caption{Overview of the Proposed Genetic Algorithm-based Method}
    \label{fig:myapproach}
\end{figure}

The generated inputs are executed with various Systems Under Test (SUTs), comprising nine open-source Java libraries designed to process structured data. This process observes how the SUTs handle these inputs. Figure \ref{fig:myapproach} represents a structured approach for generating and testing JSON files using an evolutionary algorithm. The goal is to identify unexpected outcomes, or exceptions, which are then compiled into Coverage Reports and Exception Reports. These reports measure the extent of code exercised by the inputs (e.g. instruction, line, branch, method, and class coverage) and document errors encountered during testing.

\subsection{Initial Input Generation}
The initial population of test files is generated using a JSON grammar as input. This grammar consists of a set of production rules, which are randomly selected during input generation. The structure of each JSON file is expanded based on these randomly chosen rules, allowing for high variability and structural diversity among the generated inputs. In contrast to approaches \cite{eberlein2020evolutionary,soremekun2020inputs} which begin with five manually crafted seed inputs to infer a probabilistic grammar, GA approach directly leverages random rule expansion from the base grammar. By relying solely on random selection from the original grammar, our approach naturally produces a diverse initial population without the need for prior samples. For illustration purposes, consider the following simplified grammar representing JSON generation:
\begin{verbatim}
    Grammar:
    json     → "{" pairs "}" | "{" "}"
    pairs    → pair | pair "," pairs
    pair     → string ":" value
    value    → string | number | json | "true" | "false" | "null"
    string   → "\"" letters "\""
    letters  → letter | letter letters
    letter   → "a" | "b" | ... | "Z" | digit
    number   → digit | digit number
    digit    → "0" | "1" | ... | "9"

    Possible Outputs:
    {}
    { "a": "true" }
\end{verbatim}

This grammar highlights how random expansion can lead to structurally varied JSON, demonstrating the effectiveness of the method in generating diverse initial inputs.

\subsection{Genetic Operators}\label{subsec3.1}

The evolutionary process applies genetic operators to enhance diversity and refine the quality of test inputs. This study specifically focuses on the one-point crossover and reordering mutation techniques.

\subsubsection{Crossover Mechanism}\label{subsubsec3.1.1}
Crossover mechanisms across various studies share the common goal of combining traits from parent inputs to generate diverse offspring, particularly in approaches that leverage grammar-based input generation to ensure syntactic validity. For instance, IFuzzer \cite{veggalam2016ifuzzer} grammar-aware crossover tailored to specific domains, such as interpreters, ensuring syntactic correctness, while EvoGFuzz \cite{eberlein2020evolutionary}, notably rely solely on mutation, omitting crossover altogether.

\begin{figure}[h]
    \centering
    \includegraphics[width=0.9\textwidth]{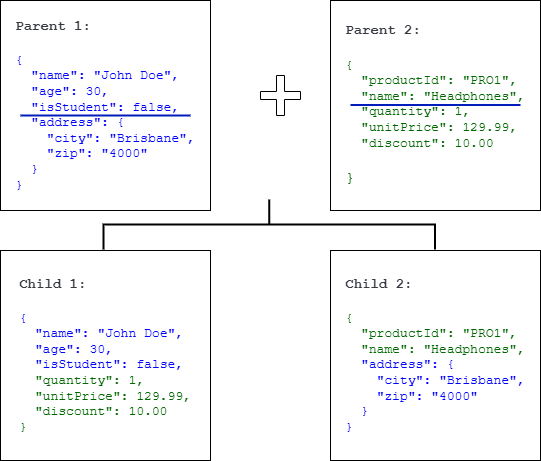}
    \caption{Example of one-point crossover on JSON inputs}
    \label{fig:crossover_example}
\end{figure}

In this study, one-point crossover is used, where a random crossover point is selected in both parent inputs, and the segments beyond this point are swapped to produce two offspring as illustrated in Figure \ref{fig:crossover_example}. This method effectively maintains the structural validity of JSON inputs while fostering meaningful diversity in test cases. When compared to other techniques like two-point crossover and uniform crossover, one-point crossover demonstrated superior performance. It is hypothesized that alternative methods may disrupt key structural relationships, or linkages, within JSON inputs, thereby reducing their effectiveness. By preserving these linkages, one-point crossover ensures that structural dependencies are maintained, enabling the generation of syntactically and semantically valid inputs that are more effective for testing. By adopting a crossover technique that integrates well with grammar-based input generation, our GA approach achieves a balance between exploration and the preservation of input validity, enhancing its ability to uncover new execution paths and vulnerabilities in the SUT.

\subsubsection{Mutation Mechanism}\label{subsubsec3.1.2}
\begin{figure}[h]
    \centering
    \includegraphics[width=0.9\textwidth]{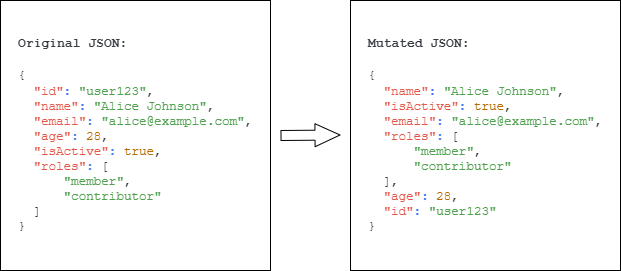}
    \caption{Example of reorder mutation on JSON input}
    \label{fig:mutation_example}
\end{figure}

Mutation strategies in evolutionary approaches introduce structural changes to test inputs while maintaining their validity. IFuzzer performs mutations by selecting random code fragments from the input and replacing them with fragments from a predefined pool corresponding to the same non-terminal, ensuring syntactic correctness \cite{veggalam2016ifuzzer}. In contrast, EvoGFuzz mutates its learned probabilistic grammar by modifying the probabilities of individual production rules, indirectly influencing the diversity of generated inputs \cite{eberlein2020evolutionary}. In this study, element reordering is used as the primary mutation technique as shown in Figure \ref{fig:mutation_example}. This method rearranges elements within JSON inputs, preserving their structural integrity while fostering diversity. Alternative mutation strategies, such as deleting elements or introducing nested structures, were also explored. However, reordering elements performed best, likely due to its ability to maintain the structural validity of the JSON files while still promoting diversity.

The choice of one-point crossover and reordering mutation as genetic operators was based on their proven ability to balance exploration and exploitation effectively. Together, these techniques maintain the structural integrity of test inputs while enhancing the algorithm's capability to uncover unique execution paths and identify potential vulnerabilities. Mutation introduces randomness, generating diverse inputs that reveal varied behaviors in the SUT, while crossover facilitates the recombination of successful traits, broadening genetic diversity. 

\subsection{Fitness Function}\label{subsec3.2}

The fitness of each test input is a critical component of the evolutionary process, as it determines which inputs are selected for subsequent generations. Initially, a weighted multi-objective fitness function is employed that incorporated both structural complexity and feedback. The weighted design enabled the algorithm to balance these objectives effectively. However, when multi-objective fitness functions are not weight-based, the objective values remain unnormalized, potentially leading to imbalanced optimization \cite{veggalam2016ifuzzer,eberlein2020evolutionary}. For example, objectives with larger numerical ranges might dominate the selection process, skewing the algorithm's focus and diminishing the overall effectiveness.

In the latest iteration of our approach, the fitness evaluation has been streamlined to focus solely on branch coverage, a metric that measures the extent to which decision branches within the code are executed during testing. This fitness score is computed as:

\begin{equation}
\text{Fitness Score}(x) = \frac{B_{\text{exec}}(x)}{B_{\text{total}}} \times 100
\end{equation}

where:
\begin{itemize}
\item x is a test input.
\item \( B_{\text{exec}}(x) \) is the number of unique branches exercised by input \( x \).
\item \( B_{\text{total}} \) is the total number of branches in the SUT.
\end{itemize}

This fitness function contributes to our study in the following ways:
\begin{itemize}
\item \textbf{Exploration:} Maximizing branch coverage encourages discovery of new execution paths, ensuring diverse code regions are exercised.
\item \textbf{Exploitation:} Inputs that reach previously uncovered branches are retained and refined, focusing on high-potential areas.
\item \textbf{Vulnerability exposure:} Deeply nested branches are often more error-prone; exercising these increases the likelihood of triggering latent vulnerabilities.
\item \textbf{Simplicity and scalability:} The single-objective nature simplifies fitness evaluation while remaining computationally efficient for large, complex SUTs.
\end{itemize}

Inputs with higher fitness scores are more likely to be selected as parents, ensuring that their traits are preserved and refined in subsequent generations. The transition to a single-objective fitness function not only simplifies the evolutionary process but also aligns it more closely with the goal of maximizing software robustness and security. Additionally, this approach reduces computational cost, making the algorithm more efficient and scalable for large and complex systems.

\subsection{Selection}\label{subsec3.3}

High-fitness inputs are selected using a tournament selection strategy, which pits a subset of candidates against each other, allowing only the most fit individuals to be chosen as parents. This method ensures that the evolutionary process is guided by high-quality inputs while maintaining genetic diversity within the population. Once selected, the parents undergo genetic operations such as crossover and mutation. Through this iterative process, the test suite evolves progressively, improving its capability to uncover critical vulnerabilities and defects by optimizing coverage in the software.

\section{Evaluation}\label{sec4}
In this study, the new GA has been evaluated and compared against the benchmark method-EvoGFuzz \cite{eberlein2020evolutionary}. The evaluation focuses on the effectiveness of both methods in terms of their ability to uncover vulnerabilities and achieve high coverage of the SUT. The metrics used for comparison include coverage reports and the types of exceptions triggered during the testing process.

\subsection{Benchmark Problems}\label{subsec4.1}

The benchmark problems were selected based on their extensive use in other research studies for evaluation purposes \cite{soremekun2020inputs,eberlein2020evolutionary}. The chosen libraries accept JSON as input, with the majority serving as serialization/deserialization libraries to convert Java objects into JSON and back. These include libraries such as Gson, Fastjson, and Minimal-json, which are commonly used for tasks like JSON parsing and serialization. Additionally, json-simple is utilized for encoding and decoding JSON text, while json-flattener is specifically designed to flatten nested JSON objects and unflatten them back. The evaluation setup involves testing the generated inputs on nine open-source Java libraries, each serving as an SUT. The open-source libraries evaluated in this study are enumerated in Table \ref{tab:sut}.

\begin{table}
\caption{Open-source projects used in the evaluation}
\label{tab:sut}
\centering
\begin{tabular}{lll}
\hline\noalign{\smallskip}
\textbf{Input Format} & \textbf{Benchmark Problem} & \textbf{Lines of Code} \\
\noalign{\smallskip}\hline\noalign{\smallskip}
\multirow{9}{*}{JSON} 
    & Argo 5.4 \cite{sourceforgeArgo}            & 8,265 \\
    & fastjson 1.2.51 \cite{fastjson}            & 166,761 \\
    & Genson 1.4 \cite{genson}                   & 18,780 \\
    & Gson 2.8.5 \cite{githubGitHubGooglegson}   & 25,172 \\
    & json-flattener 0.6.0 \cite{jsonflattener}  & 1,522 \\
    & JSONJava 20180130 \cite{JSONjava}          & 3,742 \\
    & json-simple 1.1.1 \cite{jsonsimple}        & 2,432 \\
    & MinimalJson 0.9.5 \cite{minimaljson}       & 6,350 \\
    & Pojo 0.5.1 \cite{jsonschema2pojo}          & 18,492 \\
\noalign{\smallskip}\hline
\end{tabular}
\end{table}

\subsection{Benchmark Method - Comparative Analysis}\label{subsec4.2}

For our study, the benchmark method we selected is the "Evolutionary Grammar-Based Fuzzing (EvoGFuzz)" introduced by Eberlein et al., which provides an effective mechanism for balancing both the diversity and complexity of test inputs \cite{eberlein2020evolutionary}. This \textbf{GA-based approach} is particularly well-suited for our research due to its ability to handle large and complex input spaces, which aligns with our goal of generating diverse and complex JSON inputs. EvoGFuzz builds on an existing probabilistic grammar-based fuzzer by incorporating an evolutionary algorithm into the process,  which is an extension to the work on "Inputs from Hell" by Soremekun et al. \cite{soremekun2020inputs}.

EvoGFuzz represents a significant advancement in the field of software testing, particularly in leveraging evolutionary algorithms for grammar-based input generation. Its ability to produce syntactically valid inputs minimizes errors due to invalid test cases. However, while EvoGFuzz has demonstrated notable success, certain aspects of its approach require further consideration. For instance, its reliance on probabilistic grammar can introduce biases in input generation, potentially limiting the diversity of generated inputs. Furthermore, the absence of genetic operators like crossover restricts its ability to explore novel combinations of input structures, which are often critical for uncovering deeper execution paths. These considerations highlight areas where further enhancements could make the methodology even more robust and effective.

EvoGFuzz adopts a probabilistic mechanism to input generation, leveraging grammar-based constraints to ensure that the generated inputs adhere to syntactical correctness. By assigning probabilities to different grammar rules, EvoGFuzz can guide the input generation process to favor specific patterns or structures. While this probabilistic mechanism effectively minimizes invalid inputs and aligns with the SUT's requirements, it can inadvertently introduce biases that limit the exploration of diverse or less common input structures, potentially leaving certain execution paths untested. In contrast, our study generates inputs directly from a predefined JSON grammar without relying on probabilistic methods. By randomly selecting elements based on the grammar's rules, each potential structure has an equal chance of being generated. This ensures a broader and more unbiased exploration of the input space, increasing the likelihood of uncovering edge cases and complex execution paths.

EvoGFuzz does not employ a crossover mechanism as part of its test input generation process. Instead, it focuses on generating inputs using a probabilistic grammar and relies heavily on mutation and probabilistic adjustments to the grammar to introduce variability. Specifically, EvoGFuzz alters the learned probabilistic grammar rather than mutating the test inputs directly. However, the absence of a crossover mechanism in EvoGFuzz imposes certain limitations. Without crossover, the algorithm relies primarily on mutation, which may restrict the generated inputs to variations around a smaller set of initial seeds. This can reduce the diversity of test cases and result in a narrower exploration of the input space, potentially overlooking certain vulnerabilities or performance issues. In genetic algorithms (GAs), crossover and mutation play complementary roles: crossover is instrumental for both exploration and exploitation, while mutation primarily enhances exploration \cite{tang2007memetic,vcrepinvsek2013exploration}.

In contrast, our study integrates a one-point crossover mechanism, enabling the algorithm to escape local optima by generating offspring that inherit diverse and advantageous characteristics. When mutation coupled with crossover, the approach not only broadens the genetic diversity but also refines existing input structures to maximize their fitness. Together, these genetic operators enable the evolutionary process to systematically explore deeper paths in the code, improving the likelihood of uncovering critical vulnerabilities and performance issues.

EvoGFuzz employs a fitness function that combines structural complexity and execution feedback to evaluate test inputs. This method ensures that inputs are syntactically valid while prioritizing those that trigger exceptions. However, when an input triggers an exception, its structural complexity becomes irrelevant due to the overwhelming influence of the feedback score. This bias toward simpler, fault-revealing inputs can lead to a narrower exploration of the input space, potentially overlooking structurally complex inputs that may expose deeper vulnerabilities. In contrast, the proposed approach initially employed a weight-based fitness function to balance multiple objectives, such as structural complexity and feedback. This was later refined to focus solely on branch coverage, prioritizing inputs that explore untested execution paths. By emphasizing branch coverage, the current study ensures systematic exploration, increasing the likelihood of uncovering vulnerabilities across the codebase. 

\subsection{Experimental Setup}\label{subsec4.3}
All experiments have been performed on a system running Ubuntu 20.04.6 LTS featuring an AMD Ryzen 32-Core Processor with 251 GB of RAM. Each experiment was executed 30 times, with a runtime of 600 seconds per experiment. This time frame was chosen to maintain a fair comparison with EvoGFuzz, which operates under the same conditions. The implementation was carried out using Visual Studio Code (VS Code) as the development environment. The experimental setup requires Python 3.7 and the NumPy package to run the test generation framework.

Table \ref{tab:experiment_setup} outlines the detailed setup of the experiment, showcasing different configurations for each experiment, including initial input sources, fitness functions, population size, crossover and mutation techniques applied. The experiments summarized here were designed to explore the impact of different evolutionary strategies and configurations on software vulnerability detection. Experiment 1 serves as the baseline and uses the benchmark method without applying crossover, relying solely on probabilistic grammar and mutation. This experiment is a replication of the benchmark method used for comparison. Subsequent experiments introduce variations to evaluate the impact of different configurations on performance. The experiments 2 to 4 employed probabilistic grammar derived from classic samples, focusing on a multi-objective fitness function that balanced feedback and structure scores. In these experiments, different weight distributions (equal, feedback-heavy, and structure-heavy) were tested to assess their influence on input generation. No crossover was applied, and mutation was limited to modifying the probability of production rules within the grammar.

Experiments 5 to 7 introduced more sophisticated techniques to further enhance input diversity and exploration. Crossover, specifically the one-point technique, was integrated into the process. In addition, mutation shifted from modifying the grammar to directly modifying the JSON files. The input generation method also evolved: starting with probabilistic grammar-based input generation, it later switched to random generation from a well-defined JSON grammar.

\begin{table}
\caption{Experiment setup for different configurations}
\label{tab:experiment_setup}
\centering
\begin{tabular}{lllll}
\hline\noalign{\smallskip}
\rotatebox{90}{\textbf{Exp.}} & 
\rotatebox{90}{\shortstack{\textbf{Initial} \\ \textbf{Input}}} & 
\rotatebox{90}{\shortstack{\textbf{Fitness} \\ \textbf{Function}}} & 
\rotatebox{90}{\textbf{Crossover}} & 
\rotatebox{90}{\textbf{Mutation}} \\
\noalign{\smallskip}\hline\noalign{\smallskip}
%Exp. 1 & Probabilistic grammar from classic samples & Feedback \& Structure Score & \xmark & Probability mutation in production rule \\
1 & Probabilistic grammar  & Feedback \& Structure & \xmark & Probability mutation  \\
    & from classic samples & Score &    & in production rule \\
2 & Probabilistic grammar  & Feedback \& Structure  & \xmark & Probability mutation  \\
    & from classic samples & Score (0.5, 0.5) &     & in production rule \\
3 & Probabilistic grammar & Feedback \& Structure & \xmark & Probability mutation  \\
    & from classic samples & Score (0.9, 0.1)  &    & in production rule \\
4 & Probabilistic grammar & Feedback \& Structure & \xmark & Probability mutation  \\
    & from classic samples & Score (0.1, 0.9)  &    & in production rule \\
5 & Probabilistic grammar & Branch Coverage & \cmark & \xmark  \\
    & from classic samples & Only  &    &  \\
6 & Probabilistic grammar & Branch Coverage & \cmark & Reordering elements  \\
    & from classic samples & Only  &    & in JSON files     \\
7 & Randomly generated & Branch Coverage & \cmark & Reordering elements  \\
    & from JSON grammar & Only  &    & in JSON files     \\
\noalign{\smallskip}\hline
\end{tabular}
\end{table}

\subsection{Evaluation Metrics}\label{subsec4.4}
The evaluation of both methods was based on several key metrics, including:

\begin{itemize}
    \item \textbf{Coverage:} These metrics reflect the extent to which the generated inputs exercise different parts of the SUT’s code. Higher coverage indicates a more thorough exploration of the codebase, increasing the likelihood of uncovering potential vulnerabilities or defects. Coverage is a widely used metric in software testing research as it provides a quantitative measure of how comprehensively the inputs interact with the SUT.

    \item \textbf{Exceptions Triggered:} This metric captures the types of exceptions encountered during testing, which can indicate potential vulnerabilities or areas of instability within the SUT. Tracking exceptions provides direct insight into the ability of the generated inputs to expose faults. Exception-triggering capability is also commonly employed in software testing studies to evaluate the effectiveness of test input generation methods.
\end{itemize}

These evaluation metrics were selected because they provide a comprehensive assessment of the generated inputs' effectiveness in exploring the codebase and identifying vulnerabilities. Additionally, both metrics are widely adopted in research studies within the software testing domain, ensuring comparability and relevance to existing literature \cite{eberlein2020evolutionary,soremekun2020inputs,wang2021cmfuzz,wen2023evaluating}.

\subsection{Results and Analysis}\label{subsec4.5}

The analysis of results focuses on evaluating the effectiveness of the new GA approach through two primary aspects: coverage metrics and exception types. In terms of coverage metrics, the experiments assess how well the different configurations perform across multiple benchmarks, highlighting areas where coverage improvements are observed. The exception types section provides insights into the kind of exceptions triggered during the experiments, along with the frequency and consistency of their occurrence. The 100 JSON test files generated from each experimental configuration were executed on the benchmark problems, producing two types of results: exceptions triggered and a coverage report.

\subsubsection{Coverage Metrics}\label{subsubsec4.5.1}

The results compare the performance of the benchmark method with the weighted fitness function, the new GA approach utilizing probabilistic input generation, and the final implementation of the improved evolutionary algorithm-based method across various experimental configurations. This analysis evaluates the effectiveness of each method by examining their code coverage metrics and exception-triggering capabilities.

\begin{sidewaystable}
\centering
\caption{Benchmark vs Weighted Fitness Function}
\label{tab:weight}
\small
\setlength{\tabcolsep}{3pt}
\begin{tabular}{@{}ll *{5}{ccc} @{}}
\toprule
\textbf{SUT} & \textbf{Approach} & 
\multicolumn{3}{c}{\textbf{Class}} & 
\multicolumn{3}{c}{\textbf{Method}} & 
\multicolumn{3}{c}{\textbf{Line}} & 
\multicolumn{3}{c}{\textbf{Instruction}} & 
\multicolumn{3}{c}{\textbf{Branch}} \\
\cmidrule(r){3-5} \cmidrule(lr){6-8} \cmidrule(lr){9-11} \cmidrule(lr){12-14} \cmidrule(l){15-17}
& & Max & Mean & SD & Max & Mean & SD & Max & Mean & SD & Max & Mean & SD & Max & Mean & SD \\
\midrule
Argo         & Benchmark Method & 57.89 & 31.78 & 6.64 & 35.60 & 16.80 & 4.65 & 42.49 & 20.13 & 5.85 & 37.81 & 16.44 & 4.38 & 35.91 & 15.60 & 5.66 \\
             & Experiment 2     & 57.89 & 31.86 & 7.00 & 35.92 & 16.87 & 4.88 & 42.68 & 20.27 & 6.26 & 38.24 & 16.55 & 4.73 & 36.10 & 15.99 & 6.14 \\
             & Experiment 3     & 57.89 & 31.56 & 7.00 & 36.08 & 16.68 & 4.89 & 43.80 & 20.02 & 6.25 & 39.07 & 16.35 & 4.72 & 36.65 & 15.71 & 6.10 \\
             & Experiment 4     & 57.89 & \textbf{34.43} & 6.84 & 35.76 & \textbf{18.59} & 4.76 & 42.87 & \textbf{22.49} & 6.41 & 38.48 & \textbf{18.19} & 4.88 & 36.28 & \textbf{18.18} & 6.46 \\ \hline
             
FastJson     & Benchmark Method & 26.36 & 8.15  & 2.11 & 10.80 & 3.72  & 0.91 & 5.46  & 1.77  & 0.53 & 7.33  & 1.54  & 0.63 & 2.84  & 1.17  & 0.40 \\
             & Experiment 2     & 26.36 & 8.18  & 2.11 & 10.75 & 3.73  & 0.90 & 5.40  & 1.78  & 0.54 & 7.25  & 1.56  & 0.63 & 2.74  & 1.18  & 0.40 \\
             & Experiment 3     & 26.36 & 8.19  & 2.11 & 10.75 & 3.74  & 0.90 & 5.46  & 1.79  & 0.53 & 7.32  & 1.56  & 0.63 & 2.84  & 1.19  & 0.40 \\
             & Experiment 4     & 26.36 & \textbf{8.21}  & 2.11 & 10.89 & \textbf{3.76}  & 0.90 & 5.52  & \textbf{1.80}  & 0.53 & 7.40  & \textbf{1.57}  & 0.63 & 2.88  & \textbf{1.20}  & 0.40 \\ \hline
             
Genson       & Benchmark Method & 19.03 & 6.63  & 1.31 & 9.88  & 3.89  & 0.71 & 15.87  & 5.19  & 1.29 & 8.13  & 2.41  & 0.67 & 6.76  & 1.61  & 0.58 \\
             & Experiment 2     & 20.35 & 7.13  & 1.46 & 10.35 & 4.27  & 0.85 & 17.11  & 6.03  & 1.64 & 8.68  & 3.02  & 0.91 & 7.47  & 2.23  & 0.84 \\
             & Experiment 3     & 19.91 & \textbf{7.16}  & 1.48 & 10.48 & \textbf{4.31}  & 0.85 & 17.25  & \textbf{6.10}  & 1.64 & 8.84  & \textbf{3.07}  & 0.90 & 7.68  & \textbf{2.27}  & 0.84 \\
             & Experiment 4     & 20.35 & 7.15  & 1.48 & 10.41 & 4.30  & 0.86 & 16.85  & 6.08  & 1.65 & 8.65  & 3.06  & 0.91 & 7.53  & \textbf{2.27}  & 0.85 \\ \hline
             
Gson         & Benchmark Method & 33.33 & 15.00 & 3.52 & 22.67 & 11.17 & 3.14 & 24.55 & 11.55 & 3.05 & 21.76 & 9.68  & 2.52 & 19.82 & 8.38  & 2.53 \\
             & Experiment 2     & 33.33 & 14.99 & 3.51 & 22.67 & 11.17 & 3.14 & 24.85 & 11.54 & 3.05 & 21.82 & 9.67  & 2.52 & 20.39 & 8.36  & 2.53 \\
             & Experiment 3     & 33.33 & \textbf{15.00} & 3.50 & 22.67 & 11.18 & 3.16 & 24.37 & 11.53 & 3.06 & 21.51 & 9.66  & 2.52 & 19.62 & 8.34  & 2.53 \\
             & Experiment 4     & 33.33 & 14.98 & 3.57 & 22.56 & \textbf{11.20} & 3.18 & 24.48 & \textbf{11.61} & 3.08 & 21.63 & \textbf{9.73}  & 2.54 & 19.86 & \textbf{8.44}  & 2.53 \\ \hline
             
JsonFlattener & Benchmark Method & 46.15 & \textbf{37.11} &  6.84 & 61.86 & \textbf{32.35} &  7.61 & 70.27 & \textbf{28.02} &  6.79 & 61.12 & \textbf{21.07} & 5.40 & 68.67 & 18.92 & 6.48 \\
              & Experiment 2     & 46.15 & 34.51 & 10.45 & 62.89 & 30.23 & 11.81 & 71.21 & 26.88 & 10.60 & 61.73 & 20.19 & 8.32 & 71.08 & 19.88 & 9.64 \\
              & Experiment 3     & 46.15 & 34.73 & 10.16 & 62.89 & 30.44 & 11.50 & 69.13 & 27.03 & 10.33 & 61.22 & 20.29 & 8.12 & 68.67 & \textbf{19.96} & 9.43 \\
              & Experiment 4     & 46.15 & 33.54 & 11.13 & 62.89 & 29.14 & 12.61 & 69.70 & 25.96 & 11.30 & 61.16 & 19.46 & 8.86 & 69.08 & 19.34 & 10.17\\ \hline
JSONJava     & Benchmark Method & 42.86 & \textbf{25.28} & 11.18 & 20.21 & \textbf{9.79}  & 3.87 & 16.36 & \textbf{6.55}  & 2.52 & 15.56 & \textbf{5.77}  & 2.36 & 14.19 & \textbf{4.34}  & 2.01 \\
             & Experiment 2     & 42.86 & 24.97 & 11.20 & 19.51 & 9.71  & 3.87 & 16.04 & 6.52  & 2.51 & 15.45 & 5.74  & 2.34 & 13.93 & 4.33  & 1.99 \\
             & Experiment 3     & 42.86 & 25.15 & 11.22 & 19.16 & 9.77  & 3.88 & 15.59 & 6.54  & 2.52 & 14.65 & \textbf{5.77}  & 2.36 & 12.87 & \textbf{4.34}  & 2.00 \\
             & Experiment 4     & 42.86 & 25.11 & 11.16 & 19.51 & 9.72  & 3.86 & 16.40 & 6.49  & 2.50 & 15.62 & 5.71  & 2.34 & 13.73 & 4.29  & 1.98 \\ \hline
             
Json-simple  & Benchmark Method & 75.00 & 61.20 & 6.07  & 43.75 & 27.55 & 4.94 & 55.50 & \textbf{30.43} & 6.50 & 78.16 & \textbf{16.84} & 6.23 & 49.45 & \textbf{26.38} & 7.40 \\
             & Experiment 2     & 75.00 & \textbf{61.21} & 6.24  & 43.75 & 27.52 & 5.00 & 55.13 & \textbf{30.43} & 6.60 & 77.82 & 16.83 & 6.25 & 48.63 & 26.34 & 7.49 \\
             & Experiment 3     & 75.00 & 61.20 & 6.12  & 43.75 & \textbf{27.55} & 4.95 & 55.88 & 30.42 & 6.52 & 78.16 & 16.82 & 6.23 & 48.90 & 26.33 & 7.40\\
             & Experiment 4     & 75.00 & 61.04 & 6.05  & 43.75 & 27.45 & 4.93 & 56.38 & 30.32 & 6.53 & 78.63 & 16.77 & 6.25 & 51.10 & 26.16 & 7.41 \\ \hline
             
MinimalJson  & Benchmark Method & 81.82 & 59.51 & 7.56  & 41.07 & 23.86 & 5.82 & 50.72 & 27.79 & 7.46 & 45.58 & 23.06 & 5.86 & 35.96 & 17.99 & 5.62 \\
             & Experiment 2     & 81.82 & 59.73 & 7.54  & 41.07 & 23.99 & 5.80 & 49.50 & 27.96 & 7.42 & 44.45 & 23.19 & 5.84 & 36.80 & 17.99 & 5.57 \\
             & Experiment 3     & 81.82 & \textbf{59.80} & 7.58  & 41.07 & \textbf{24.05} & 5.82 & 50.17 & \textbf{28.03} & 7.45 & 45.07 & \textbf{23.26} & 5.86 & 36.24 & 18.06 & 5.59 \\
             & Experiment 4     & 81.82 & 59.73 & 7.53  & 41.43 & 24.04 & 5.81 & 50.17 & 28.02 & 7.44 & 44.69 & 23.24 & 5.86 & 36.24 & \textbf{18.12} & 5.63 \\ \hline
             
Pojo         & Benchmark Method & 42.67 & 24.28 & 4.11  & 35.67 & 15.56 & 6.14 & 30.78 & 10.64 & 5.85 & 28.26 & 8.59 & 5.64 & 17.76 & 5.40 & 3.96 \\
             & Experiment 2     & 42.67 & 24.65 & 5.02  & 36.11 & 16.33 & 7.02 & 30.82 & 11.40 & 6.14 & 28.11 & 9.40 & 5.83 & 17.96 & 6.06 & 4.18 \\
             & Experiment 3     & 42.67 & \textbf{24.71} & 4.96  & 36.11 & \textbf{16.43} & 6.98 & 30.96 & \textbf{11.49} & 6.16 & 28.25 & \textbf{9.48} & 5.86 & 18.06 & 6.11 & 4.18 \\
             & Experiment 4     & 42.67 & 24.62 & 5.25  & 36.11 & 16.35 & 7.32 & 31.41 & 11.48 & 6.36 & 28.66 & 9.50 & 6.02 & 17.96 & \textbf{6.13} & 4.31 \\ 
\bottomrule
\end{tabular}
\end{sidewaystable}

Table \ref{tab:weight} presents the coverage results comparing the weighted fitness function with the benchmark method. It highlights metrics such as class, method, line, instruction, and branch coverage. Each experiment’s maximum, mean, and standard deviation (SD) values are provided, allowing for an in-depth comparison of the performance across different configurations. The coverage values reported are not cumulative; instead, they represent the average coverage achieved by a single test input when executed on the benchmark problems.

Experiments 2, 3, and 4 represent the application of a weighted multi-objective fitness function combining both feedback score (ability to trigger exceptions) and structure score (complexity of input). These experiments apply different weight configurations to explore the trade-offs between these objectives:
\begin{itemize}
    \item Experiment 2: Equal weight distribution of (0.5, 0.5) between feedback and structure.
    \item Experiment 3: Prioritizes feedback score with weights (0.9, 0.1).
    \item Experiment 4: Prioritizes structure score with weights (0.1, 0.9).
\end{itemize}

The bolded values indicate the cases where the performance of a method stood out, either by achieving the highest coverage. The results indicate that the weighted fitness function generally outperforms the benchmark method across all coverage metrics, with the exception of JsonFlattener and JSONJava. Most SUTs showed measurable gains with the introduction of weighted objectives, reflecting the effectiveness of balancing feedback and structural complexity. Notably, Experiment 3, which assigns greater weight to the feedback score (0.9 feedback, 0.1 structure), achieved consistently higher coverage across several metrics. This configuration proved particularly effective at guiding the search toward inputs that trigger exceptions and expose vulnerabilities by prioritizing meaningful execution paths. Overall, emphasizing feedback over structural complexity facilitated the exploration of fault-prone code regions and contributed to improved test effectiveness. However, the choice of weight configuration should be guided by the specific objectives of the study. For example, if the goal is to trigger exceptions and explore runtime behavior, a feedback-focused weighting is ideal; whereas, for studies targeting structural diversity or grammar conformance, greater emphasis on input complexity may be beneficial.

\begin{sidewaystable}
\centering
\caption{Benchmark vs New Evolutionary Algorithm}
\label{tab:cross_mutation}
\small
\setlength{\tabcolsep}{3pt}
\begin{tabular}{@{}ll *{5}{ccc} @{}}
\toprule
\textbf{SUT} & \textbf{Approach} & 
\multicolumn{3}{c}{\textbf{Class}} & 
\multicolumn{3}{c}{\textbf{Method}} & 
\multicolumn{3}{c}{\textbf{Line}} & 
\multicolumn{3}{c}{\textbf{Instruction}} & 
\multicolumn{3}{c}{\textbf{Branch}} \\
\cmidrule(r){3-5} \cmidrule(lr){6-8} \cmidrule(lr){9-11} \cmidrule(lr){12-14} \cmidrule(l){15-17}
& & Max & Mean & SD & Max & Mean & SD & Max & Mean & SD & Max & Mean & SD & Max & Mean & SD \\
\midrule
Argo         & Benchmark Method & 57.89 & 31.78 & 6.64 & 35.60 & 16.80 & 4.65 & 42.49 & 20.13 & 5.85 & 37.81 & 16.44 & 4.38 & 35.91 & 15.60 & 5.66 \\
             & Experiment 5     & 57.89 & \textbf{46.37} & 1.52 & 34.79 & \textbf{27.60} & 0.97 & 43.11 & \textbf{37.16} & 1.35 & 38.32 & \textbf{29.02} & 1.27 & 36.46 & \textbf{35.09} & 1.74 \\
             & Experiment 6     & 57.89 & 46.21 & 1.74 & 34.79 & 27.49 & 1.11 & 43.11 & 37.03 & 1.60 & 38.32 & 28.91 & 1.44 & 36.46 & 34.95 & 2.09 \\ \hline
FastJson     & Benchmark Method & 26.36 & 8.15  & 2.11 & 10.80 & 3.72  & 0.91 & 5.46  & 1.77  & 0.53 & 7.33  & 1.54  & 0.63 & 2.84  & 1.17  & 0.40 \\
             & Experiment 5     & 26.36 & \textbf{10.57} & 1.62 & 10.84 & \textbf{5.14}  & 0.60 & 5.78  & 3.14  & 0.31 & 7.59  & 2.72  & 0.50 & 3.17  & 2.44  & 0.18 \\
             & Experiment 6     & 26.36 & \textbf{10.57} & 1.62 & 10.84 & \textbf{5.14}  & 0.60 & 5.85  & \textbf{3.16}  & 0.32 & 7.65  & \textbf{2.73}  & 0.51 & 3.24  & \textbf{2.45}  & 0.19 \\ \hline
Genson       & Benchmark Method & 19.03 & 6.36  & 1.31 & 9.88  & 3.89  & 0.71 & 15.87  & 5.19  & 1.29 & 8.13  & 2.41  & 0.67 & 6.76  & 1.61  & 0.58 \\
             & Experiment 5     & 19.91 & \textbf{8.06}  & 1.11 & 9.88  & \textbf{5.50}  & 0.46 & 16.22 & \textbf{8.73}  & 0.79 & 8.31  & \textbf{4.60}  & 0.41 & 7.07  & \textbf{3.72}  & 0.37 \\
             & Experiment 6     & 19.47 & \textbf{8.06}  & 1.11 & 9.88  & \textbf{5.50}  & 0.46 & 16.22 & 8.71  & 0.80 & 8.31  & 4.58  & 0.42 & 7.07  & 3.71  & 0.39 \\ \hline
Gson         & Benchmark Method & 33.33 & 15.00 & 3.52 & 22.67 & 11.17 & 3.14 & 24.55 & 11.55 & 3.05 & 21.76 & 9.68 & 2.52 & 19.82 & 8.38 & 2.53 \\
             & Experiment 5     & 33.33 & \textbf{19.24} & 1.48 & 22.67 & \textbf{16.17} & 0.79 & 24.79 & \textbf{18.45} & 0.85 & 21.88 & \textbf{15.38} & 0.80 & 20.59 & \textbf{14.89} & 0.83 \\
             & Experiment 6     & 33.33 & 19.23 & 1.48 & 22.67 & \textbf{16.17} & 0.79 & 24.79 & 18.42 & 0.87 & 21.88 & 15.34 & 0.82 & 20.55 & 14.83 & 0.86 \\ \hline
JsonFlattener & Benchmark Method & 46.15 & 37.11 & 6.84 & 61.86 & 32.35 & 7.61 & 70.27 & 28.02 & 6.79 & 61.12 & 21.07 & 5.40 & 68.67 & 18.92 & 6.48 \\
              & Experiment 5     & 46.15 & 46.04 & 1.52 & 59.79 & 58.05 & 3.66 & 72.54 & 66.93 & 4.78 & 63.87 & 58.08 & 4.88 & 72.29 & 66.23 & 5.03 \\
              & Experiment 6     & 46.15 & \textbf{46.07} & 1.43 & 59.79 & \textbf{58.71} & 3.21 & 72.73 & \textbf{67.92} & 4.51 & 63.87 & \textbf{59.26} & 4.44 & 73.09 & \textbf{66.77} & 5.31 \\ \hline
JSONJava     & Benchmark Method & 42.86 & 25.28 & 11.18 & 20.21 & 9.79  & 3.87 & 16.36  & 6.55  & 2.52 & 15.56 & 5.77  & 2.36 & 14.19 & 4.34  & 2.01 \\
             & Experiment 5     & 42.86 & \textbf{42.82} & 0.84  & 17.77 & 16.86 & 0.39 & 16.90 & 15.50 & 0.68 & 14.96 & 13.73 & 0.60 & 12.48 & 11.72 & 0.55 \\
             & Experiment 6     & 42.86 & 42.78 & 0.96  & 17.77 & \textbf{16.95} & 0.44 & 16.95 & \textbf{15.61} & 0.82 & 15.00 & \textbf{13.85} & 0.73 & 12.67 & \textbf{11.78} & 0.71 \\ \hline
Json-simple  & Benchmark Method & 75.00 & 61.20 & 6.07 & 43.75 & 27.55 & 4.94 & 55.50 & 30.43 & 6.50 & 78.16 & 16.84 & 6.23 & 49.45 & 26.38 & 7.40 \\
             & Experiment 5     & 75.00 & 74.79 & 1.64 & 43.75 & 35.39 & 1.20 & 58.25 & 49.85 & 2.00 & 79.20 & 26.31 & 5.31 & 51.65 & 46.21 & 1.89 \\
             & Experiment 6     & 75.00 & \textbf{74.95} & 0.87 & 43.75 & \textbf{35.47} & 0.98 & 58.25 & \textbf{49.98} & 1.82 & 79.20 & \textbf{26.39} & 5.30 & 51.65 & \textbf{46.27} & 1.76 \\ \hline
MinimalJson  & Benchmark Method & 81.82 & 59.51 & 7.56 & 41.07 & 23.86 & 5.82 & 50.72 & 27.79 & 7.46 & 45.58 & 23.06 & 5.86 & 35.96 & 17.99 & 5.62 \\
             & Experiment 5     & 81.82 & \textbf{77.19} & 1.26 & 41.07 & \textbf{37.69} & 0.97 & 51.50 & 47.12 & 1.31 & 46.64 & 38.64 & 1.21 & 35.96 & 34.32 & 1.20 \\
             & Experiment 6     & 81.82 & 77.18 & 1.30 & 41.07 & 37.61 & 1.06 & 51.50 & \textbf{47.22} & 1.54 & 46.64 & \textbf{38.75} & 1.37 & 35.96 & \textbf{34.46} & 1.51 \\ \hline
Pojo         & Benchmark Method & 42.67 & 24.28 & 4.11  & 35.67 & 15.56 & 6.14  & 30.78 & 10.64 & 5.85  & 28.26 & 8.59 & 5.64  & 17.76 & 5.40  & 3.96 \\
             & Experiment 5     & 42.67 & \textbf{37.27} & 1.05  & 36.11 & \textbf{32.33} & 1.30  & 31.54 & \textbf{26.31} & 1.43  & 28.87 & \textbf{23.57} & 1.36  & 18.47 & \textbf{17.56} & 1.01 \\
             & Experiment 6     & 42.67 & 37.23 & 1.26  & 36.11 & 32.24 & 1.54  & 31.54 & 26.22 & 1.57  & 28.87 & 23.51 & 1.49  & 18.37 & 17.40 & 1.14 \\ 
\bottomrule
\end{tabular}
\end{sidewaystable}

The results presented in Table \ref{tab:cross_mutation}, compares the benchmark method against the GA approach incorporating crossover (Experiment 5) and both crossover and mutation (Experiment 6). The findings demonstrate that our approach significantly outperforms the benchmark method across most coverage metrics.

Substantial improvements were observed in other SUTs, particularly Argo, JsonFlattener, JSONJava, and Json-simple, where GA-based approach achieved significantly higher coverage metrics. The introduction of one-point crossover in Experiment 5 produced marked improvements over the benchmark method. For instance, in JSONJava and Json-simple, the mean branch coverage reached 66.77\% and 49.49\%, respectively, confirming the effectiveness of crossover in exploring new input spaces and increasing coverage.

The combination of crossover and mutation in Experiment 6 provided further improvements for certain SUTs. In JsonFlattener, Gson, and Pojo, the addition of mutation slightly boosted the mean branch coverage. However, in some cases, such as Argo and Json-simple, the results were comparable to or slightly lower than Experiment 5.

Overall, the results in Table \ref{tab:cross_mutation} highlight the superiority of the new evolutionary algorithm. The inclusion of genetic operators, particularly crossover, enhances the exploration and exploitation of the input space, increasing the likelihood of discovering unique execution paths and software vulnerabilities. Mutation further contributes to input diversity by introducing structural randomness, enabling the discovery of previously untested paths.

\begin{sidewaystable}
\centering
\caption{Benchmark vs Finalized Evolutionary Algorithm}
\label{tab:results_final}
\small
\setlength{\tabcolsep}{3pt}
\begin{tabular}{@{}ll *{5}{ccc} @{}}
\toprule
\textbf{SUT} & \textbf{Approach} & 
\multicolumn{3}{c}{\textbf{Class}} & 
\multicolumn{3}{c}{\textbf{Method}} & 
\multicolumn{3}{c}{\textbf{Line}} & 
\multicolumn{3}{c}{\textbf{Instruction}} & 
\multicolumn{3}{c}{\textbf{Branch}} \\
\cmidrule(r){3-5} \cmidrule(lr){6-8} \cmidrule(lr){9-11} \cmidrule(lr){12-14} \cmidrule(l){15-17}
& & Max & Mean & SD & Max & Mean & SD & Max & Mean & SD & Max & Mean & SD & Max & Mean & SD \\
\midrule
Argo         & Benchmark Method & 57.89 & 31.78 & 6.64 & 35.60 & 16.80 & 4.65 & 42.49 & 20.13 & 5.85 & 37.81 & 16.44 & 4.38 & 35.91 & 15.60 & 5.66 \\ 
             & Experiment 5     & 57.89 & 46.37 & 1.52 & 34.79 & 27.60 & 0.97 & 43.11 & 37.16 & 1.35 & 38.32 & 29.02 & 1.27 & 36.46 & 35.09 & 1.74 \\ 
             & Experiment 6     & 57.89 & 46.21 & 1.74 & 34.79 & 27.49 & 1.11 & 43.11 & 37.03 & 1.60 & 38.32 & 28.91 & 1.44 & 36.46 & 34.95 & 2.09 \\ 
             & Experiment 7     & 57.89 & \textbf{52.36} & 1.57 & 36.41 & \textbf{29.23} & 1.09 & 46.53 & \textbf{40.64} & 1.38 & 40.63 & \textbf{31.36} & 1.28 & 40.70 & \textbf{39.76} & 1.57 \\ \hline
             
FastJson     & Benchmark Method & 26.36 & 8.15  & 2.11 & 10.80 & 3.72  & 0.91 & 5.46  & 1.77  & 0.53 & 7.33  & 1.54  & 0.63 & 2.84  & 1.17  & 0.40 \\ 
             & Experiment 5     & 26.36 & 10.57 & 1.62 & 10.84 & 5.14  & 0.60 & 5.78  & 3.14  & 0.31 & 7.59  & 2.72  & 0.50 & 3.17  & 2.44  & 0.18 \\ 
             & Experiment 6     & 26.36 & 10.57 & 1.62 & 10.84 & 5.14  & 0.60 & 5.85  & 3.16  & 0.32 & 7.65  & 2.73  & 0.51 & 3.24  & 2.45  & 0.19 \\ 
             & Experiment 7     & 26.36 & \textbf{10.94} & 1.65 & 10.99 & \textbf{5.30}  & 0.64 & 6.15  & \textbf{3.48}  & 0.34 & 8.02  & \textbf{3.08}  & 0.52 & 3.56  & \textbf{2.79}  & 0.20 \\ \hline
             
Genson       & Benchmark Method & 19.03 & 6.36  & 1.31 & 9.88   & 3.89  & 0.71 & 15.87 & 5.19  & 1.29 & 8.13  & 2.41  & 0.67 & 6.76  & 1.61  & 0.58 \\ 
             & Experiment 5     & 19.91 & 8.06  & 1.11 & 9.88   & 5.50  & 0.46 & 16.22 & 8.73  & 0.79 & 8.31  & 4.60  & 0.41 & 7.07  & 3.72  & 0.37 \\ 
             & Experiment 6     & 19.47 & 8.06  & 1.11 & 9.88   & 5.50  & 0.46 & 16.22 & 8.71  & 0.80 & 8.31  & 4.58  & 0.42 & 7.07  & 3.71  & 0.39 \\ 
             & Experiment 7     & 20.35 & \textbf{8.66}  & 1.16 & 10.55  & \textbf{5.97}  & 0.54 & 17.88 & \textbf{10.09} & 0.94 & 9.17  & \textbf{5.32}  & 0.47 & 8.35  & \textbf{4.76}  & 0.47 \\ \hline

Gson         & Benchmark Method & 33.33 & 15.00 & 3.52 & 22.67 & 11.17 & 3.14 & 24.55 & 11.55 & 3.05 & 21.76 & 9.68  & 2.52 & 19.82 & 8.38  & 2.53 \\ 
             & Experiment 5     & 33.33 & \textbf{19.24} & 1.48 & 22.67 & 16.17 & 0.79 & 24.79 & 18.45 & 0.85 & 21.88 & 15.38 & 0.80 & 20.59 & 14.89 & 0.83 \\ 
             & Experiment 6     & 33.33 & 19.23 & 1.48 & 22.67 & 16.17 & 0.79 & 24.79 & 18.42 & 0.87 & 21.88 & 15.34 & 0.82 & 20.55 & 14.83 & 0.86 \\ 
             & Experiment 7     & 33.95 & 19.18 & 1.67 & 22.90 & \textbf{16.37} & 0.88 & 26.38 & \textbf{19.99} & 0.85 & 23.53 & \textbf{16.97} & 0.81 & 22.91 & \textbf{17.29} & 0.80 \\ \hline
             
JsonFlattener & Benchmark Method & 46.15 & 37.11 & 6.84 & 61.86 & 32.35 & 7.61 & 70.27 & 28.02 & 6.79 & 61.12 & 21.07 & 5.40 & 68.67 & 18.92 & 6.48 \\ 
              & Experiment 5     & 46.15 & 46.04 & 1.52 & 59.79 & 58.05 & 3.66 & 72.54 & 66.93 & 4.78 & 63.87 & 58.08 & 4.88 & 72.29 & 66.23 & 5.03 \\ 
              & Experiment 6     & 46.15 & \textbf{46.07} & 1.43 & 59.79 & \textbf{58.71} & 3.21 & 72.73 & \textbf{67.92} & 4.51 & 63.87 & \textbf{59.26} & 4.44 & 73.09 & \textbf{66.77} & 5.31 \\ 
              & Experiment 7     & 46.15 & 45.83 & 2.93 & 59.79 & 56.14 & 6.21 & 70.27 & 63.33 & 8.90 & 62.19 & 54.54 & 9.13 & 71.89 & 62.40 & 9.67 \\ \hline
              
JSONJava     & Benchmark Method & 42.86 & 25.28 & 11.18 & 20.21 & 9.79  & 3.87 & 16.36 & 6.55  & 2.52 & 15.56 & 5.77  & 2.36 & 14.19 & 4.34  & 2.01 \\ 
             & Experiment 5     & 42.86 & \textbf{42.82} & 0.84  & 17.77 & 16.86 & 0.39 & 16.90 & 15.50 & 0.68 & 14.96 & 13.73 & 0.60 & 12.48 & 11.72 & 0.55 \\ 
             & Experiment 6     & 42.86 & 42.78 & 0.96  & 17.77 & 16.95 & 0.44 & 16.95 & 15.61 & 0.82 & 15.00 & 13.85 & 0.73 & 12.67 & 11.78 & 0.71 \\ 
             & Experiment 7     & 42.86 & 42.79 & 1.21  & 18.47 & \textbf{17.74} & 0.49 & 18.21 & \textbf{17.46} & 0.66 & 16.42 & \textbf{15.73} & 0.63 & 15.05 & \textbf{14.60} & 0.66 \\ \hline
             
Json-simple  & Benchmark Method & 75.00 & 61.20 & 6.07 & 43.75 & 27.55 & 4.49 & 55.50 & 30.43 & 6.50 & 78.16 & 16.84 & 6.23 & 49.45 & 26.38 & 7.40 \\ 
             & Experiment 5     & 75.00 & 74.79 & 1.64 & 43.75 & 35.39 & 1.20 & 58.25 & 49.85 & 2.00 & 79.20 & 26.31 & 5.31 & 51.65 & 46.21 & 1.89 \\ 
             & Experiment 6     & 75.00 & \textbf{74.95} & 0.87 & 43.75 & \textbf{35.47} & 0.98 & 58.25 & 49.98 & 1.82 & 79.20 & 26.39 & 5.30 & 51.65 & 46.27 & 1.76 \\ 
             & Experiment 7     & 75.00 & 74.82 & 1.62 & 43.75 & 35.40 & 1.29 & 60.25 & \textbf{53.13} & 2.15 & 80.68 & \textbf{28.37} & 5.33 & 56.04 & \textbf{51.57} & 2.12 \\ \hline
             
MinimalJson  & Benchmark Method & 81.82 & 59.51 & 7.56 & 41.07 & 23.86 & 5.82 & 50.72 & 27.79 & 7.46 & 45.58 & 23.06 & 5.86 & 35.96 & 17.99 & 5.62 \\ 
             & Experiment 5     & 81.82 & \textbf{77.19} & 1.26 & 41.07 & 37.69 & 0.97 & 51.50 & 47.12 & 1.31 & 46.64 & 38.64 & 1.21 & 35.96 & 34.32 & 1.20 \\ 
             & Experiment 6     & 81.82 & 77.18 & 1.30 & 41.07 & 37.61 & 1.06 & 51.50 & 47.22 & 1.54 & 46.64 & 38.75 & 1.37 & 35.96 & 34.46 & 1.51 \\ 
             & Experiment 7     & 81.82 & 77.17 & 1.48 & 41.43 & \textbf{38.15} & 1.00 & 53.28 & \textbf{49.51} & 1.34 & 48.82 & \textbf{41.43} & 1.26 & 42.13 & \textbf{41.16} & 1.40 \\ \hline
             
Pojo         & Benchmark Method & 42.67 & 24.28 & 4.11 & 35.67 & 15.56 & 6.14 & 30.78 & 10.64 & 5.85 & 28.26 & 8.59  & 5.64 & 17.76 & 5.40  & 3.96 \\ 
             & Experiment 5     & 42.67 & \textbf{37.27} & 1.05 & 36.11 & \textbf{32.33} & 1.30 & 31.54 & \textbf{26.31} & 1.43 & 28.87 & \textbf{23.57} & 1.36 & 18.47 & \textbf{17.56}  & 1.01 \\ 
             & Experiment 6     & 42.67 & 37.23 & 1.26 & 36.11 & 32.24 & 1.54 & 31.54 & 26.22 & 1.57 & 28.87 & 23.51 & 1.49 & 18.37 & 17.40& 1.14 \\ 
             & Experiment 7     & 42.67 & 36.54 & 2.38 & 36.54 & 31.44 & 3.14 & 31.90 & 24.89 & 3.19 & 29.08 & 22.29 & 3.07 & 19.07 & 16.65 & 2.20 \\      
\bottomrule
\end{tabular}
\end{sidewaystable}

The coverage metrics of the benchmark method, the evolutionary algorithm with probabilistic input generation (Experiments 5 and 6), and the final evolutionary algorithm with random input generation (Experiment 7) are compared in Table \ref{tab:results_final}. Both Experiments 6 and 7 implement the proposed genetic algorithm-based approach, incorporating genetic operators such as crossover and mutation. The primary difference between these two configurations lies in their input generation strategy: Experiment 6 leverages probabilistic grammar learned from initial samples, while Experiment 7 directly employs random input generation from a well-defined JSON grammar.

Across nearly all coverage metrics, Experiment 7 consistently outperforms both the benchmark and probabilistic grammar setups. The advantage of random generation stems from its unbiased nature, where each production rule in the grammar has an equal probability of being selected. This uniform randomness facilitates broader structural diversity, allowing the evolutionary process to explore input regions that may be entirely missed in probabilistic approaches, which tend to favor frequently observed patterns from the initial training samples. As a result, random generation helps prevent premature convergence and promotes exploration of rarely exercised program paths.

The experimental results demonstrate substantial improvements over the benchmark method. On average, class coverage increased by 39.8\%, with strong performance in JSONJava and MinimalJson. Method coverage improved by 62.4\%, particularly in Argo and Genson. Even greater gains were seen in line coverage (105.0\%) and instruction coverage (114.0\%), with JsonFlattener and Genson showing notable advances. The most significant improvement occurred in branch coverage, which rose by 166.0\%, underscoring the effectiveness of the approach in exploring complex conditional logic and uncovering deeper execution paths. These findings validate the robustness of the evolutionary algorithm in achieving comprehensive and high-quality software testing.

\subsubsection{Exceptions Triggered}\label{subsubsec4.5.3}

\begin{sidewaystable}

\centering
\caption{Exception Detected Across Benchmark Problems}
\label{tab:exception}
\begin{tabular}{lllcccc}
\hline\noalign{\smallskip}

\shortstack{\textbf{Benchmark} \\ \textbf{Problem}} &
\textbf{Exception Types} &
\textbf{Location} &
\textbf{EvoGFuzz} &
\shortstack{\textbf{Weight} \\ \textbf{(0.5, 0.5)}} &
\shortstack{\textbf{Weight} \\ \textbf{(0.9, 0.1)}} &
\shortstack{\textbf{Weight} \\ \textbf{(0.1, 0.9)}} \\

\noalign{\smallskip}\hline\noalign{\smallskip}
Argo                       & InvalidSyntaxException   & argo.saj.InvalidSyntaxRuntime & 30 & 30 & 30 & 30 \\ 
                            &                           & Exception nS1:41                                &    &    &    &     \\
                           &                          & argo.saj.InvalidSyntaxRuntime & 27 & 30 & 30 & 26 \\ 
                           &                           & Exception nS1:50                                &    &    &    &     \\
                           &                          & argo.saj.InvalidSyntaxRuntime & 30 & 30 & 30 & 30 \\ 
                           &                           & Exception nS1:60                                &    &    &    &     \\\hline
Genson                     & NullPointerException     & com.owlike.genson.stream.    & 30 & 30 & 30 & 30 \\ 
&                           & JsonWriter:414                            &    &    &    &     \\\hline
JsonFlattener              & ClassCastException       & com.github.wmnameless.json.base.            & 30 & 30 & 30 & 30 \\ 
                           &                          & JacksonJsonValue:74                         &    &    &    &    \\ 
                           &                          & com.github.wmnameless.json.base.            & 30 & 30 & 30 & 30 \\ 
                           &                          & JacksonJsonValue:79                         &    &    &    &    \\ 
                           & NullPointerException     & com.github.wmnameless.json.     & 30 & 30 & 30 & 30 \\ 
                           &                          & unflattener.JsonUnflattener:532                         &    &    &    &    \\ 
                           & RuntimeException         & com.github.wmnameless.json.base.            & 30 & 30 & 30 & 30 \\ 
                           &                          & JacksonJsonCore:49                          &    &    &    &    \\ \hline
Pojo                       & StringIndexOutOf & org.jsonschema2pojo.util.       & 30 & 30 & 30 & 30 \\ 
                           & BoundsException                                 & NameHelper:46                   &    &    &    &    \\
\noalign{\smallskip}\hline
\end{tabular}
\end{sidewaystable}

This section provides an analysis of the exception types triggered in the benchmark problems and identifies the specific locations where these exceptions occurred under different experimental configurations.

Out of the 9 benchmark problems tested, 4 triggered exceptions: Argo, Genson, JsonFlattener, and Pojo. Table \ref{tab:exception} summarizes the exception types triggered in each benchmark, along with their occurrence locations. Each benchmark revealed distinct exception types, providing valuable insights into their underlying issues.

The Argo benchmark triggered the \textit{argo.saj.InvalidSyntaxException}, indicating parsing issues in the input data. In Genson, the \textit{java.lang.NullPointerEx-ception} was consistently triggered, reflecting potential problems with null reference handling. JsonFlattener exhibited multiple exceptions, including \textit{java.lang-.NullPointerException}, \textit{java.lang.ClassCastException}, and \textit{java.lang.Runtime-Exception}, highlighting type casting errors and general runtime issues during the JSON flattening process. In Pojo, the \textit{java.lang.StringIndexOutOfBounds-Exception} occurred, pointing to problems with string handling during data transformation.

The experiment was conducted 30 times, and Table \ref{tab:exception} shows the frequency of each exception under various configurations. For instance, in Argo, the \textit{argo.saj.InvalidSyntaxException} at nS2:50 was triggered in 27 runs when using EvoGFuzz. When applying the weight-based method with equal weighting or prioritizing the feedback score, the exception was triggered in all 30 runs. However, in Experiment 4, where more weight was assigned to the structure score, the exception occurred in only 26 runs. These findings emphasize the importance of balancing fitness objectives in the multi-objective fitness function, as prioritizing specific objectives can influence the consistency of exception detection.

Based on these findings, it is evident that deciding the appropriate weights between multiple objectives in the fitness function is crucial to optimizing performance and maximizing exception detection.

All the experiments successfully triggered all the exceptions listed in the Table \ref{tab:exception}.

\section{Conclusion and Future Work}\label{sec5}

This study introduces a novel GA-based framework for software vulnerability detection, showcasing the effectiveness of adaptive, feedback-driven learning in guiding input generation. By leveraging execution feedback from the SUT, the evolutionary process dynamically refines test inputs over successive generations, enabling targeted exploration of previously untested execution paths. The use of genetic operators plays a pivotal role in this adaptive search: crossover facilitates exploration by fundamentally altering input structures and generating entirely new combinations of test features, thus navigating a broader search space; in contrast, mutation provides finer-grained adjustments, allowing localized exploitation around promising solutions. This balance between exploration and exploitation ensures both diversity and depth in test input generation. Furthermore, by shifting from probabilistic input generation to random derivation from a well-defined JSON grammar, the approach eliminates inherent biases and guarantees comprehensive coverage of complex input structures. Experimental results demonstrate consistent improvements across all key coverage metrics, underscoring the robustness and generalizability of the proposed framework.

Building on the strengths of the proposed GA-based method, future work can focus on broadening its applicability and enhancing its robustness. As modern software systems grow more complex, with increasingly intricate logic and interactions, improving the algorithm’s ability to handle highly dynamic and deeply nested execution paths will make it even more effective and broadly applicable. Additionally, exploring adaptive control over the genetic operators could further help maintain a strong balance between exploring new input spaces and refining promising solutions throughout the evolutionary process. These directions represent natural progressions of the current work, aimed at maximizing its impact in advancing secure and reliable software systems.

\bibliography{sn-bibliography}% common bib file

\end{document}